\documentclass[twocolumn,showpacs,preprintnumbers,amsmath,amssymb,prl]{revtex4-1}
\usepackage{stackengine,graphicx} 
\usepackage{epsfig} 
\usepackage{dcolumn} 
\usepackage{bm} 
\usepackage{xcolor}
\usepackage{xspace}
\usepackage{comment}
\usepackage{ulem}
\usepackage{pdfpages}
\graphicspath{{./}{Figures//}}
\newcounter{saveenumi}

\newcommand{\be}{\begin{enumerate}}
\newcommand{\ee}{\end{enumerate}}

\definecolor{robgreen}{rgb}{0.333,0.62,0.18}
\definecolor{linebrown}{HTML}{bd6e00}
\definecolor{lineblue}{HTML}{0000ff}
\definecolor{linegreen}{HTML}{217200}
\definecolor{linered}{HTML}{d20d0d}

\newcommand{\RAA}{\AA$^{-1}$}

\def\fd3m{Fd$\overline 3$m}
\def\p1bar{P$\overline 1$}
\def\i41amd{I4$_{1}$/amd}
\def\t2g{$t_{2g}$}

\def\rangez{$(0 \leq z \leq 2)$}
\def\kfeses{K$_x$Fe$_{2-y}$Se$_{2-z}$S$_z$}

\newcommand{\qmax}{\ensuremath{Q_{\mathrm{max}}}\xspace}
\newcommand{\qmin}{\ensuremath{Q_{\mathrm{min}}}\xspace}

\newcommand{\pdfgui}{\textsc{PDFgui}\xspace}

\newcommand{\pdfgetn}{\textsc{PDFgetN}\xspace}

\DeclareRobustCommand{\rout}[1]{\textcolor{red}{\vphantom{#1}}}
\DeclareRobustCommand{\routtwo}[1]{\textcolor{red}{\vphantom{#1}}}
\newcommand{\ins}[1]{\textcolor{black}{#1}}
\newcommand{\instwo}[1]{\textcolor{black}{#1}}

\makeatletter
\AtBeginDocument{\let\LS@rot\@undefined}
\makeatother
\begin{document}
%
%
\begin{abstract}
A detailed account of the local atomic structure and disorder at 5~K across the phase diagram of the high temperature superconductor \kfeses\ \rangez\ is obtained from neutron total scattering and associated atomic pair distribution function (PDF) approaches.
Various model independent and model dependent aspects of the analysis reveal a high level of structural complexity on the nanometer length-scale.
Evidence is found for considerable disorder in the $c$-axis stacking of the FeSe$_{1-x}$S$_{x}$ \rout{slabs with no observable signs}\ins{slabs without observable signs} of turbostratic character of the disorder.
In contrast to the related FeCh (Ch = S, Se) type superconductors, substantial Fe-vacancies are present in \kfeses, deemed detrimental for superconductivity when ordered.
Our study suggests that the distribution of vacancies significantly modifies the iron-chalcogen bond-length distribution, in agreement with observed evolution of the PDF signal.
A crossover like transition is observed at a composition of $z\approx1$, from a \routtwo{predominantly vacancy}\instwo{correlated} disorder state at the selenium end to a more vacancy-ordered (VO) \routtwo{phase}\instwo{state} closer to the sulfur end of the phase diagram.
The S-content dependent measures of the local structure are found to exhibit distinct behavior on either side of this crossover, correlating well with the evolution of the superconducting state to that of a magnetic semiconductor towards the $z\approx2$ end.
The behavior reinforces the idea of the intimate relationship of correlated Fe-vacancies order in the local structure and the emergent electronic properties.
\end{abstract}

\title{Correlated disorder to order crossover in the local structure of \kfeses}

\author{P.~Mangelis,$^{1}$, R.~J.~Koch,$^{2,*}$, H.~Lei,$^{2,^\dag}$, R.~B.~Neder,$^{3}$, M.~T.~McDonnell,$^{4\ddagger}$, M.~Feygenson,$^{4,\S}$, C.~Petrovic,$^{2}$, A.~Lappas,$^{1,*}$, and E.~S.~Bozin$^{2}$}

 \affiliation{$^{1}$Institute of Electronic Structure and Laser, Foundation for Research and Technology—Hellas, Vassilika Vouton, 711 10 Heraklion, Greece }
 \affiliation{$^{2}$Condensed Matter Physics and Materials Science Department, Brookhaven National Laboratory, Upton, New York~11973, USA}
 \affiliation{$^{3}$Institute of Condensed Matter Physics, Friedrich-Alexander-Universit\"at Erlangen-N\"urnberg, Staudtstr.~3, 91058 Erlangen, Germany}
 \affiliation{$^{4}$Neutron Scattering Division, Oak Ridge National Laboratory, Oak Ridge, Tennessee~37831, USA}

 \altaffiliation{rkoch@bnl.gov\\lappas@iesl.forth.gr\\
 $^\dag$ Present address: Department of Physics and Beijing Key Laboratory of Opto-electronic Functional Materials \& Micro-nano Devices, Renmin University of China, Beijing, China.\\
 $^\ddagger$ Present Address: Computer Science and Mathematics Division, Oak Ridge National Laboratory, Oak Ridge, TN~37831, USA\\
 $^\S$ Present address: Forschungszentrum J\" ulich, JCNS, D-52425~J\" ulich, Germany}

\date{\today}
\maketitle
%
\section{Introduction}
\label{sec:intro}

Recently discovered 122 type iron-based superconductors, with the formula A$_x$Fe$_2$Se$_2$ (A = alkali metals or Tl)~\cite{guo_superconductivity_2010, ye_common_2011}, and relatively high transition temperatures ($T_c\approx30$~K) as compared to the simpler FeSe-11 type counterparts ($T_c\approx8$~K)~\cite{Hsu2008}, have garnered significant interest in correlated electron systems research.
This class of unconventional superconductors combines unique properties such as the coexistence of superconductivity and long-range antiferromagnetic (AF) order with large magnetic moments and N\'eel temperatures that far exceed room temperature~\cite{bao_novel_2011, wang_antiferromagnetic_2011, yan_electronic_2012,krzton-maziopa_superconductivity_2016}, \ins{while spectroscopic studies suggest the presence of a superconducting gap below 8~meV}~\cite{charnukha_optical_2012}.
Their behavior sets the stage for a tantalizing correlated electron materials conceptual problem, somewhat contrasting the cuprates, where superconductivity emerges from the antiferromagnetic Mott insulating state.
The puzzle becomes even more complex as a wide range of ensemble average~\cite{louca_hybrid_2013, carr_structure_2014,shoemaker_phase_2012,lazarevic_vacancy-induced_2012} and local structure~\cite{ksenofontov_phase_2011, ding_influence_2013, ricci_nanoscale_2011, wang_structural_2012} methods have shown strong evidence for nanoscale phase separation, \ins{particularly perpendicular to the iron-selenide planes}~\cite{charnukha_nanoscale_2012}.

\begin{figure}[tbp]
\includegraphics[width=1.0\columnwidth]{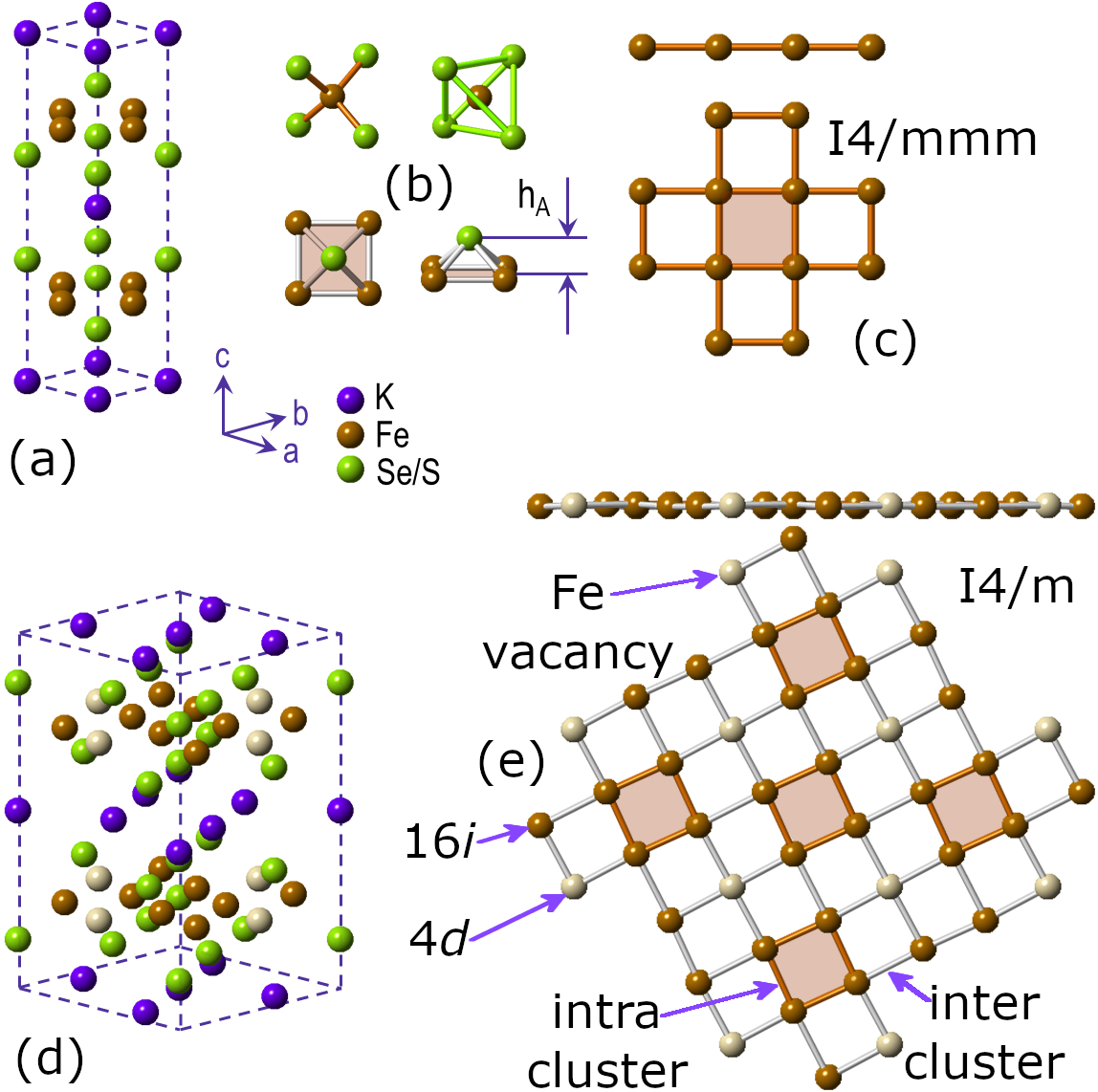}
\caption{\label{fig:fig1} (Color online) Two possible crystallographic models of the atomic structure of \kfeses.
(a) A perspective view of a single unit cell in the $I4/mmm$ Fe-vacancy-free model.
(b) A single FeCh$_4$ tetrahedron and other structural elements discussed in the text.
$h_A$ denotes the anion height.
(c) The Fe sub-lattice viewed both along the $a-b$ plane (top) and down the $c$-axis (bottom) in the $I4/mmm$ model.
(d) A perspective view of a single unit cell in the $I4/m$ VO model.
(e) The Fe sub-lattice in the $I4/m$ VO model as viewed both along the $a-b$ plane (top) and down the $c$-axis (bottom).
Symmetry-distinct Fe-sites ($4d$ and $16i$), Fe vacancies, as well as various Fe-Fe interatomic distances discussed in text are indicated by arrows.
}
\end{figure}

In an effort to describe the coexistence of superconducting and AF states, a common approach portrays a majority insulating AF Fe-VO A$_2$Fe$_4$Se$_5$ phase (space group $I4/m$), which is separated from the minority superconducting A$_x$Fe$_2$Se$_2$ phase (space group $I4/mmm$)~\cite{lei_phase_2011,texier_nmr_2012,friemel_reciprocal-space_2012}.
While the latter is characterized by fully occupied Fe atomic sites (Fig.~\ref{fig:fig1}a), the presence of Fe vacancies in the former (Fig.~\ref{fig:fig1}d) \rout{were}\ins{was} believed to be detrimental to superconductivity~\cite{li_phase_2012,friemel_conformity_2012}.
An understanding has then emerged inferring that what makes the materials non-superconducting is the magnetism bearing $\sqrt{5}\times\sqrt{5}$ long-range ordering of Fe vacancies.
This is described in the $I4/m$ tetragonal symmetry, with two crystallographically distinct Fe-atomic sites, where one ($16i$) is fully occupied, while the other ($4d$) remains empty.
Earlier neutron total scattering studies~\cite{mangelis_nanoscale_2018} of K$_x$Fe$_{2-y}$Se$_2$ and the sulfide analogue K$_x$Fe$_{2-y}$S$_2$ revealed that the two-phase description reflects a sub-nanometer length scale, rendering such a model equivalent to an Fe-vacancy-disordered $I4/m$ K$_{2-x}$Fe$_{4+y}$Ch$_{5}$ (Ch = Se, S) single-phase case.
The inhomogeneity and structural complexity of A$_x$Fe$_{2-y}$Se$_2$ makes it difficult to derive concrete conclusions about the nature of the relationship between the atomic structure and the observed physical properties.
Since magnetic and insulating $I4/m$ and superconducting $I4/mmm$ patches are mixed on the scale of about 100~nm~\cite{wang_microstructure_2011, li_phase_2012}, proximity effects of magnetic patches on superconducting islands are important~\cite{PhysRevB.80.174510}.
\ins{Indeed, spectroscopy measurements have suggested the existence of a Josephson-coupled phase in the Se end-member}~\cite{homes_optical_2012}.

In an effort to resolve the delicate interplay of structural defects and properties, in situ scanning electron microscopy studies revealed that during the cooling of a specimen to room temperature, the formation mechanism of the superconducting phase in K$_x$Fe$_{2-y}$Se$_2$ passes through an `imperfect' Fe-vacancy disorder-to-order transition, which is deemed responsible for the phase separation~\cite{liu_formation_2016}.
Crucial to mediating the physical response and the distribution of the vacancies appears to be the thermal treatment, including the quenching conditions~\cite{ding_influence_2013, wang_disordered_2015, PhysRevB.91.064513}.
Indeed, the Fe-vacancy order to disorder transition can be achieved by a simple high-temperature annealing process~\cite{song_phase_2011,wang_fe-vacancy_2018}, an approach which has also been pin-pointed by recent high energy, single-crystal X-ray diffraction experiments and Monte Carlo simulations, where superconductivity in quenched K$_x$Fe$_{2-y}$Se$_2$ crystals appears at the Fe-vacancy order-to-disorder boundary~\cite{duan_appearance_2018}.
Notably, rapid quenching in K$_x$Fe$_{2-y}$Se$_2$ leads to a giant increase in the critical current density and yields specimens of higher $T_c$~\cite{PhysRevB.84.212502}

In view of these intricate relationships, Wu et al. have suggested that the non-superconducting, alkali-modified FeSe-based magnetic insulators, which possess such Fe VO structures, should be considered as the parent compounds of the superconducting phases~\cite{wu_overview_2015}.
Disordering the Fe-vacancy order of the parent magnetic insulating phase, K$_2$Fe$_4$Se$_5$, has been proposed as crucial for the emergence of superconductivity in K$_x$Fe$_{2-y}$Se$_2$~\cite{wang_disordered_2015}.
Following this conceptual approach, X-ray absorption fine-structure studies~\cite{iadecola_large_2012, ryu_local_2012} have prodvided evidence for local disorder in the structure of superconducting K$_x$Fe$_{2-y}$Se$_2$, suggesting that a non-zero population of Fe atoms at the $4d$ site (Fig.~\ref{fig:fig1}e) is the key structural parameter for bulk superconductivity.
As the debate still goes on as of what the role of Fe vacancies might be in the emergence of superconductivity and collapse of magnetism (or vice versa), seeking good model systems where similar features can be studied ultimately helps to broaden our knowledge base.

In this endeavor, the isostructural \kfeses\ \rangez\ solid-solutions offer such a possibility as the substitution of Se by the isovalent S suppresses the superconducting state and gives rise to a magnetic semiconductor at $z \geq\ 1.6$~\cite{lei_phase_2011}.
Since the interplay of superconductivity and magnetism across the \kfeses\ series seems to rely on intriguing structural details we have employed neutron total scattering combined with atomic pair distribution function (PDF) analysis to probe the local atomic structure that may correlate with such notable changes in the behavior.
The study was carried over a dense grid of compositions at 5~K, the temperature at which superconductivity is observed in the selenium-rich part of the phase diagram, but gets suppressed towards the sulfur-rich end of the series.
Our study corroberates previous reports of disorder in the stacking of the FeSe$_{1-x}$S$_{x}$ slabs, which reflects yet another ingredient in an already structurally complex material, and has been previously linked to local potassium ordering and segregation, as well as superconductivity.
Moreover, by complementing neutron PDF analysis with simulations based on large atomistic models, we explored the subtle nanoscale changes in the interatomic distances and the evolution of vacancy distribution in the Fe-chalcogen layers in the \kfeses\ system.
We demonstrate that numerous local structural quantities correlate well with $T_c$, being marked by an Fe-vacancy \instwo{correlated} disorder to order crossover around the $z = 1$ composition. 
\ins{From this we can infer that the Fe-vacancy distribution likely plays a critical role in the electronic properties}
\rout{This serves to reinforce the idea that the Fe-vacancy distribution is the key structural parameter influencing the properties.}

\section{Methods}
\label{sec:meth}
Single crystals of \kfeses\ $(0 \leq z \leq 2)$ were grown by self-flux method, \rout{as described in detail elsewhere}\ins{and energy-dispersive x-ray spectroscopy (EDX) found no appreciable impurity elements, as described in detail elsewhere}~\cite{lei_phase_2011}.
\ins{Crystals were} pulverized into fine powders \ins{and seived to a maximum size of 40~$\mu$m.}
Samples were thoroughly characterized by X-ray powder diffraction, magnetic susceptibility, and electrical resistivity measurements, as previously reported~\cite{lei_phase_2011}.

Experiments were performed at NOMAD instrument~\cite{neuefeind_nanoscale_2012} at the Spallation Neutron Source at Oak Ridge National Laboratory.
Powdered samples were loaded in extruded vanadium containers and sealed under inert atmosphere, where helium was used as the exchange gas.
Data were collected at 5~K, then corrected and reduced using standard protocols~\cite{egami;b;utbp12}.
Neutron PDFs were obtained using the \pdfgetn\ program~\cite{peter;jac00} over a range from \qmin\ = 0.5~\RAA\ to \qmax\ = 26~\RAA.
\ins{Diffraction analysis did not find any evidence of crystal size broadining, suggesting crystal sizes were greater than 100~nm.}
Further details on these measurements are provided in Appendix~\ref{sec:appa}.

Measured and reduced data, including $I(Q)$, $F(Q)$ and $G(r)$, were analyzed using various structure-model independent approaches, outlined in Appendix~\ref{sec:appb}.
When performing structure-model dependent analysis of the same data, two structural models are relevant, with a space group of $I4/mmm$ or $I4/m$.
The details of these are summarized in Fig.~\ref{fig:fig1}, with finer details presented in Appendix~\ref{sec:appc}.
Small-box PDF refinements were carried out using one or both of these models with the \pdfgui\ software~\cite{farro;jpcm07} as described in Appendix~\ref{sec:appd}.
Large-box Simulated PDFs were computed with the DISCUS software package~\cite{Proffen1997} using a model with space group $I4/m$ as described in Appendix~\ref{sec:appe}.
\begin{figure}[tbp]
\includegraphics[width=1.0\columnwidth]{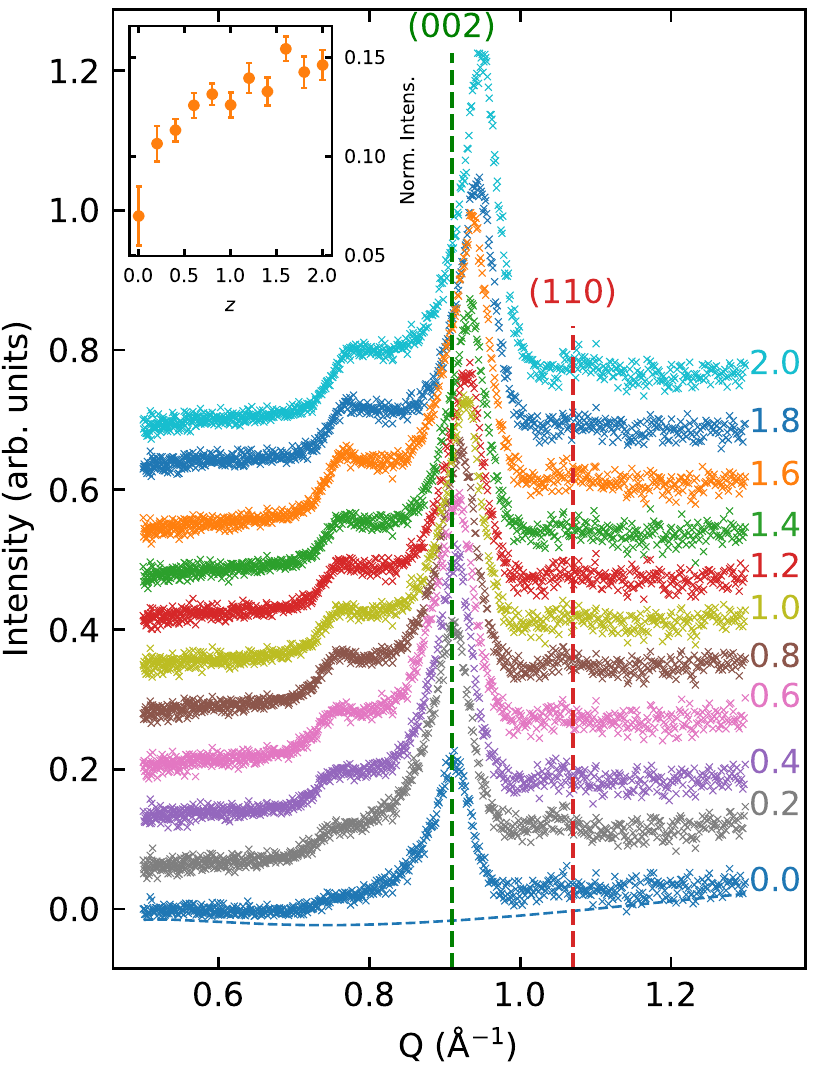}
\caption{\label{fig:fig2} (Color online) Vacancy ordering in \kfeses.
Neutron powder diffraction data (sulphur content, $z$, as indicated) in the vicinity of the (110) and (002) reflections, positions of which for $z=0$ sample, are represented by vertical red and green dashed lines, respectively.
Patterns are offset vertically by 0.07 y-axis units for clarity, and a background curve is plotted for the $z = 0$ datset to highlight the diffuse (110) feature.
The presence of a sharp (110) reflection would indicate perfect vacancy ordering (the $I4/m$ model).
Observation of broad and diffuse intensity of this reflection would imply correlated vacancy disorder.
A broad shoulder at low-$Q$ side of the (002) reflection is noticeable for each composition.
The evolution with $z$ of its normalized integrated intensity is shown in the inset.
The origin of this feature is subsequently addressed in Fig.~\ref{fig:fig3} and in the text.
}
\end{figure}

\begin{figure}[tbp]
\includegraphics[width=1.0\columnwidth]{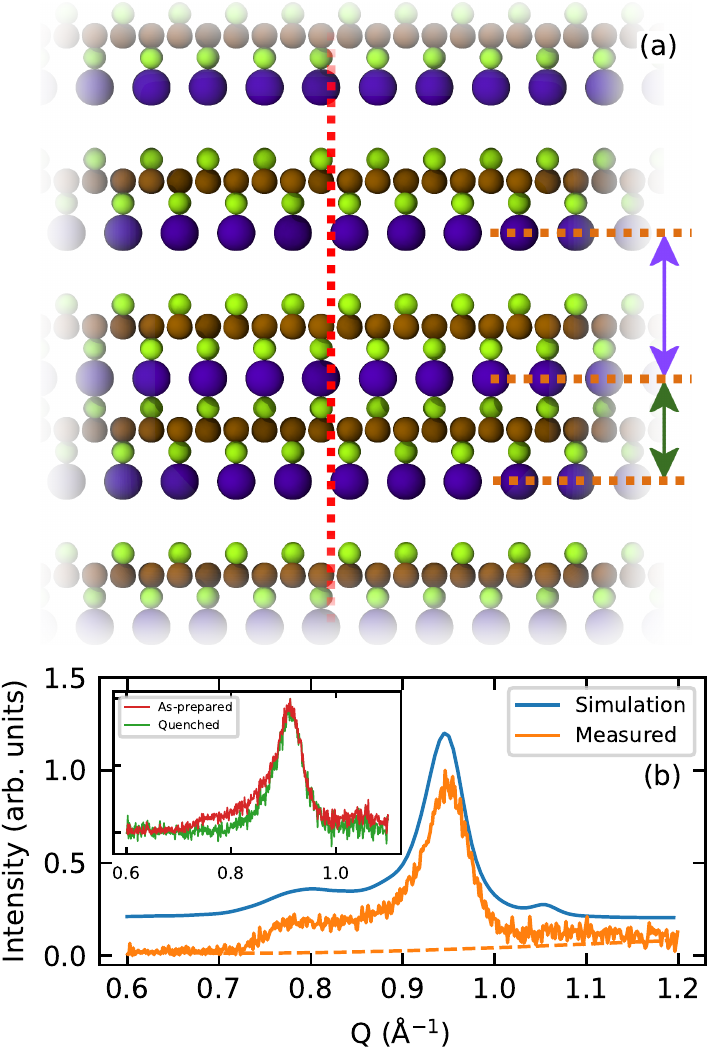}
\caption{\label{fig:fig3}(Color online) Observation of stacking disorder in \kfeses\ system.
(a) A side view of an atomistic model depicting two distinct inter-layer distances along the $c$-axis, represented by either a purple or a green double-side arrow.
Green spheres represent chalcogen, brown spheres iron, and purple spheres potassium atoms.
The red vertical dashed line highlights the lack of turbostratic disorder, as discussed in the text.
(b) The simulated neutron diffraction pattern (blue profile) of K$_{x}$Fe$_{2-y}$S$_{2}$ for the case depicted in (a), where two distinct inter-layer distances are present.
Also shown is the experimental diffraction pattern for K$_{x}$Fe$_{2-y}$S$_{2}$ sample at 5~K with the associated background (orange profile and dashed line, respectively) featuring (002) basal reflection at around 0.95~\AA$^{-1}$.
The feature at $Q\approx0.8$~\RAA\ is an irrational basal reflection implying the presence of a second, longer inter-layer spacing.
This feature is observed for all as-prepared samples studied here (Fig.~\ref{fig:fig2}).
Inset in (b) features data for the K$_{x}$Fe$_{2-y}$Se$_{2}$ sample at 5~K, for both as-prepared (red line) and thermally quenched (green line) variants.
}
\end{figure}

\section{Results and Discussion}
\label{sec:results}
\subsection{Neutron diffraction}
We start the \kfeses\ system assessment by looking at the neutron powder diffraction data as seen by the low scattering angle detector bank of the NOMAD instrument, Fig.~\ref{fig:fig2}.
The data across the entire composition series display a broad, diffuse feature at the expected location (red dashed line in Fig.~\ref{fig:fig2}) of the (110) peak associated with the $\sqrt{5}\times\sqrt{5}$ long-range ordering of Fe vacancies.
Macroscopic separation of a VO phase with a vacancy-free phase would not produce such a broad feature, but would instead produce a relatively sharp, albeit weak, (110) peak.
The diffuse nature of the observed (110) peak suggests that idealized vacancy-order is not achieved, but rather that a type of imperfect order or ``correlated disorder''~\cite{Welberry1985, Keen2015} exists across the series.
This is commonly observed in other order-disorder type systems such as, for example, in copper-gold intermetallic alloys~\cite{Cowley1965, Moss1968, Moss2005, Owen2017}.

Additionally, diffraction data across the entire composition series show a broad and bi-modal distribution of intensity at the expected location of the (002) peak, as marked by a green dashed line in Fig.~\ref{fig:fig2}.
Focusing on the specific example of K$_{x}$Fe$_{2- y}$S$_{2}$ in Fig.~\ref{fig:fig3}b, we note that the primary peak at $\sim0.95$~\RAA\ can be indexed in either the $I4/mmm$ or $I4/m$ space group as (002), but the additional low-$Q$ shoulder-like feature at $\sim0.8$~\RAA\ cannot be indexed in either space group.
Such additional diffuse intensity in the vicinity of inter-layer diffraction peaks is common in clay systems with more than one inter-layer distance~\cite{Plancon2004,Guinier1994, Drits1990}.

DISCUS simulations considering large super-cells in the K$_{x}$Fe$_{2- y}$S$_{2}$ system with two distinct inter-layer distances, 6.55~\AA\ (66~\% prevalence) and 8.0~\AA\ (34~\% prevalence) (see Fig.~\ref{fig:fig3}a) reproduce this experimentally observed bi-modal intensity distribution well, as seen in Fig.~\ref{fig:fig3}b.
Importantly, these two distinct inter-layer distances exist in the \textit{same} crystal, not segregated to distinct phases.
A phase segregated physical mixture would produce two distinct diffraction features, rather than the continuous intensity distribution seen in Fig.~\ref{fig:fig2} and Fig.~\ref{fig:fig3}b.
The relative intensity of the additional low-$Q$ shoulder-like feature increases with increasing sulfur content (Fig.~\ref{fig:fig2} inset), suggesting that the prevalence of a second longer inter-layer distances increases with increasing sulfur content.
Comparison of the data for two K$_{x}$Fe$_{2- y}$Se$_{2}$ samples obtained by different thermal treatment reveals that this type of disorder can be suppressed by quenching (see inset to Fig.~\ref{fig:fig3}(b)).

\begin{figure}[tbp]
\includegraphics[width=1.0\columnwidth]{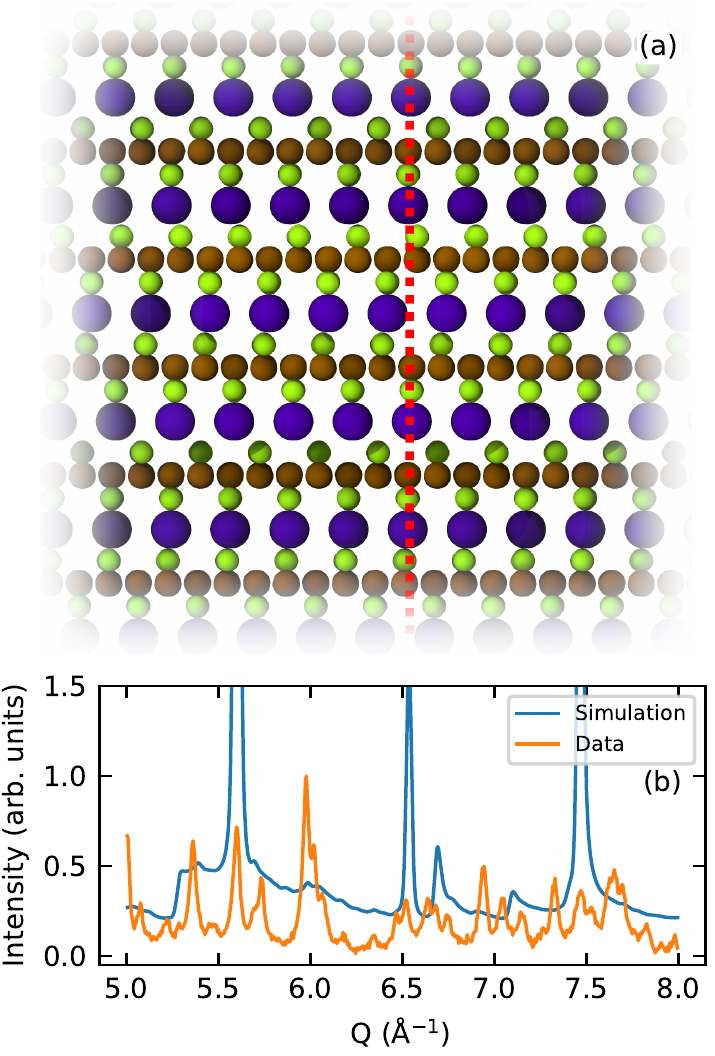}
\caption{\label{fig:fig4} (Color online) Observation of stacking disorder in \kfeses\ system.
(a) A side view of an atomistic model depicting turbostratic disorder, or uniform random layer displacements along the $a-b$ lattice directions, highlighted by the vertical red dashed line.
Atoms are represented as in Fig.~\ref{fig:fig3}, with the $c$-axis running vertically.
(b) The neutron diffraction pattern simulated for K$_{x}$Fe$_{2-y}$S$_{2}$ (blue profile) in the case of turbostratic disorder depicted in (a).
This type of disorder reduces the number of observed Bragg reflections, and produces characteristic saw-tooth like features, as discussed in the text.
The experimental data for K$_{x}$Fe$_{2-y}$S$_{2}$ (orange profile) collected at 5~K do not display the features that would imply turbostratic character of the disorder.
}
\end{figure}

Similar DISCUS simulations with a single inter-layer distance but random shifts along the $a$ and $b$ lattice directions (known as turbostratic disorder, see Fig.~\ref{fig:fig4}a) produce a diffraction pattern characteristic for such disorder, where $hkl$ peaks are broadened into indistinct $hk$-bands~\cite{Drits1990, Ufer2004a}.
As can be seen in Fig.~\ref{fig:fig4}b, this significantly reduces the number of distinct Bragg peaks, and creates a characteristic saw-tooth pattern.
In Fig.~\ref{fig:fig4}b we again present an example diffraction pattern for K$_{x}$Fe$_{2- y}$S$_{2}$, where no such saw-tooth like features are observed.

These simulations effectively demonstrate that the \kfeses\ specimens exhibit at least two distinct inter-layer distances, without observable turbostratic disorder.
This anisotropic crystallinity naturally suppresses inter-layer correlations, while intra-layer correlations persist.
Additionally, the broad nature of the (110) $\sqrt{5}\times\sqrt{5}$ ordering peak indicated that an imperfect or correlated local vacancy order is likely present, a feature which will necessarily only manifest in the local atomic structure.
For these reasons, our small-box PDF modelling has focused explicitly only on the very local structure ($r < 10$~\AA).

Such dual-interlayer distance has been reported previously, and was linked to potassium segregation and local ordering~\cite{PhysRevB.91.064513}.
This additional diffraction feature is the only evidence of such local segregation and potassium ordering found here.
No evidence of this behavior was found in the PDF studies discussed below, although the PDF in this case is not particularly sensitive to potassium.

\begin{figure}[tbp]
\includegraphics[width=1.0\columnwidth]{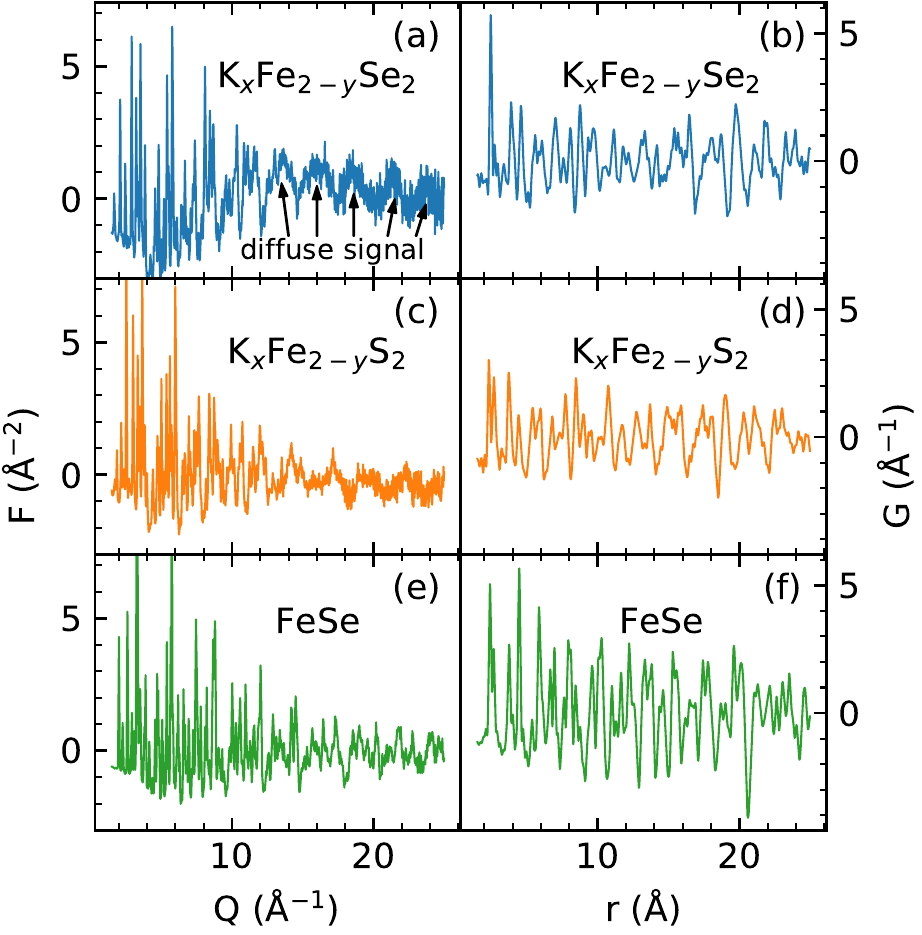}
\caption{\label{fig:fig5} (Color online) Comparison of neutron total scattering derived data for several samples of interest.
Reduced total scattering structure functions, $F(Q)$, are shown in the left panels ($1.5 \leq Q \leq 25$~\RAA), while corresponding PDFs, $G(r)$, are displayed in the right panels ($1.5 \leq r \leq 25$~\AA).
The panels feature (a), (b) K$_{x}$Fe$_{2- y}$Se$_{2}$, (c), (d) K$_{x}$Fe$_{2- y}$S$_{2}$, and (e), (f) FeSe.
Data for potassium containing samples were collected at 5~K, while for FeSe at 10~K.
}
\end{figure}

Reduced total scattering structure functions, $F(Q)$, for the K$_{x}$Fe$_{2- y}$Se$_{2}$ and K$_{x}$Fe$_{2- y}$S$_{2}$ end-members are shown in Fig.~\ref{fig:fig5}a and Fig.~\ref{fig:fig5}c, respectively, along with the related FeSe reference in Fig.~\ref{fig:fig5}e.
By inspecting these data one readily sees that Bragg peaks at high-$Q$ get suppressed when moving from the sulfur to the selenium end-member, or even when moving from the FeSe compound to either 122-type system, and that appreciable diffuse signal appears.
Suppression of Bragg peaks and appearance of diffuse intensity are indicators for the presence of disorder, suggesting that the relative amount of disorder increases when moving from the related FeSe phase to K$_{x}$Fe$_{2- y}$S$_{2}$ and then to K$_{x}$Fe$_{2- y}$Se$_{2}$.

The diffuse signal in $F(Q)$ is present in the data for the entire \kfeses\ series, as evident in Fig.~\ref{fig:fig6}a, and has a sinusoidal character at high-$Q$.
Fitting with a damped Sine function in the range $12 \leq Q \leq 25$~\RAA\ allows us to quantify the amplitude and frequency of this oscillatory behavior as a function of sulfur content $z$, shown in Fig.~\ref{fig:fig6}d and Fig.~\ref{fig:fig6}e.
We note that these oscillations are the strongest in the selenium end-member, and are gradually suppressed with increasing sulfur content, reaching a minimum in the range $1.2 \leq z \leq 1.6$.
Above $z = 1.6$, these oscillations again increase in amplitude.
This suppression and recovery is highlighted when viewing the fitted Sine functions, as can be seen in Fig.~\ref{fig:fig6}b and Fig.~\ref{fig:fig6}c.
The frequency of the wave follows a similar two-regime behavior, delineated by the composition at $z = 1.6$.

\subsection{PDF analysis}
Interpretation of these Sine-like oscillations of the diffuse signal in $Q$-space is not straightforward.
However, the presence of a sinusoidal oscillation in $F(Q)$ should map to the PDF $G(r)$, as the two are related by a Fourier transform.
Specifically, a periodic signal in $F(Q)$ with frequency $f$ and amplitude $A$ should manifest as an enhancement (proportional to $A$) of the peak in $G(r)$ at $r = 2\pi f$.

Indeed, this agrees with qualitative assessments of the PDFs.
The frequency and relative amplitude of the periodic signal in $F(Q)$ parallels the position and relative sharpness (compared to higher-$r$ peaks) of the first PDF peak.
The PDF of K$_{x}$Fe$_{2- y}$Se$_{2}$, shown in Fig~\ref{fig:fig5}b, has a sharp first peak at $r \approx 2.43$~\AA, followed by relatively broad features at higher-$r$.
Interestingly, the observed maximum of this feature is nearly twice as intense as any other feature in the PDF.
Conversely, the PDF of K$_{x}$Fe$_{2- y}$S$_{2}$, shown in Fig~\ref{fig:fig5}d, shows a peak at $r \approx 2.38$~\AA\ which is of comparable sharpness to higher-$r$ peaks.

Typically, the observation of $r$-dependent PDF peak widths is associated with correlated atomic motion~\cite{jeong;jpc99}.
Generally speaking, these relative widths describe the nature of the bonding in the material~\cite{jeong;prb03}.
For example, an ``infinitely'' sharp first PDF peak would indicate that the nearest-neighbor pairs always move in phase, and that this bond is extremely rigid.
Conversely, a first PDF peak that is of the same width as higher-$r$ peaks would indicate that nearest-neighbor pairs do not influence each-other strongly. 
The origin of this behavior in the system studied here will be addressed below.

The PDFs across the composition series at 5~K are shown in Fig.~\ref{fig:fig7}a, and the positions of a higher-$r$ peak as well as the first peak, representing Fe-Ch correlations, are shown in Fig.~\ref{fig:fig7}b and Fig.~\ref{fig:fig7}c, respectively.
Interestingly, the position of the Fe-Ch peak is nearly unchanged in the range $0.0 \leq z \leq 1.0$, signifying that the Fe-Ch pair distance is unaffected by the substitution of sulfur for selenium.
This is in stark contrast to the typical Vegard's law-type behavior, predicting a linear change in lattice parameter as a function of chemical substitution due to steric effects~\cite{vegar;zp21}.
This counter-intuitive behavior is however a strictly local effect, as the expected linear pair-distance behavior is recovered if we consider the position of most higher-$r$ peaks (Fig.~\ref{fig:fig7}a and~\ref{fig:fig7}b).
This is further evidence that the very local structure has a distinct two-regime type behavior.

\begin{figure}[tbp]
\includegraphics[width=1.0\columnwidth]{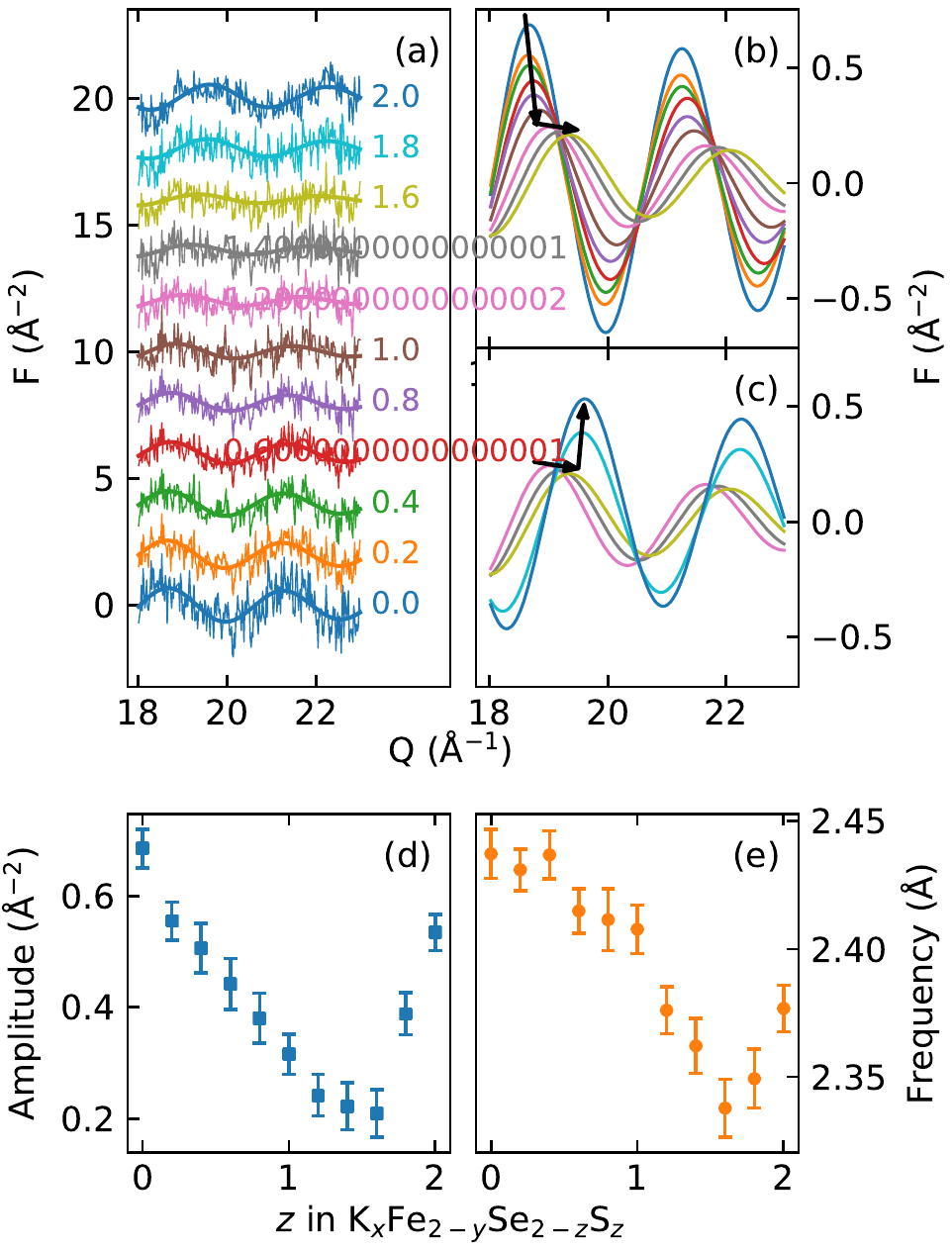}
\caption{\label{fig:fig6} (Color online) Analysis of the diffuse scattering signal in $F(Q)$ at high momentum transfers $Q$ in \kfeses.
(a) The measured (lighter line) and fitted (heavier line) $F(Q)$ signal.
The curves are sequentially offset vertically by 2~\AA$^{-2}$ for clarity.
(b) The model $F(Q)$ signal over a regime of concentrations where the oscillation amplitude and frequency decrease.
(c) The model $F(Q)$ signal over a regime of concentrations where oscillation amplitude and frequency increase.
(d) The amplitude of $F(Q)$ diffuse oscillations as a function of sulfur content, $z$.
(e) The frequency of diffuse oscillations as a function of $z$.
Colors represent sulfur content, $z$, as indicated, and in (b) and (c) are consistent with those in (a).
}
\end{figure}

\begin{figure}[tbp]
\includegraphics[width=1.0\columnwidth]{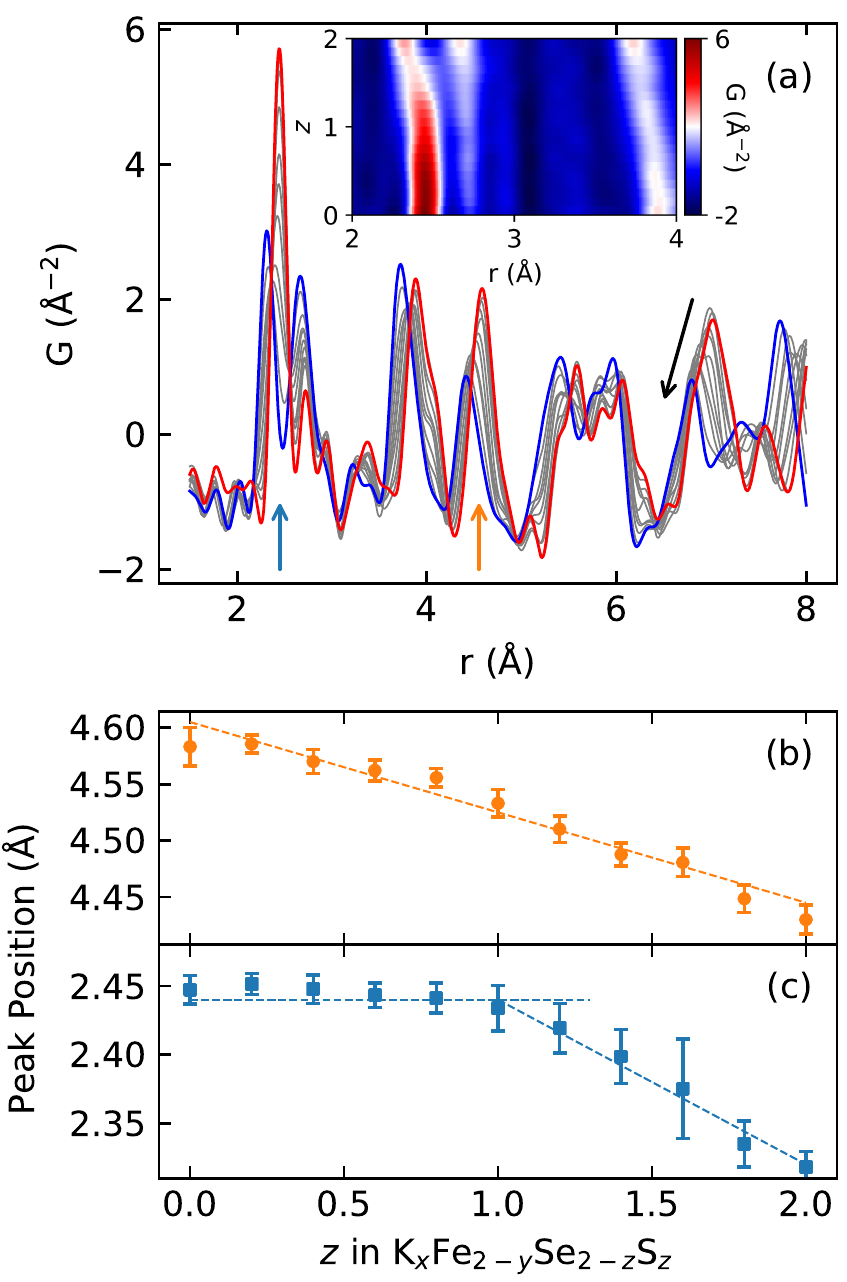}
\caption{\label{fig:fig7}(Color online) 5~K PDF data of the \kfeses\ series.
(a) The experimental PDFs in the local structure range $1.5 \leq r \leq 8$~\AA.
The sulfur and selenium end-members are shown with a blue and red line, respectively, while intermediate compositions are shown in gray.
A black arrow is included to emphasize the continuous evolution of features at high $r$.
Inset is a false color map representation of the same PDF data plotted vs $r$ and $z$, highlighting the lack of apparent change in the position of the first PDF peak at $\sim2.48$~\AA\ as a function of sulfur content until approximately $z=1$.
The peak starts evolving for higher $z$ values.
(b) A plot of the peak position of the feature marked by an orange arrow in (a) as a function of sulfur content $z$.
A dashed line is a guide to the eye, included to highlight the nearly linear behavior of this peak position with $z$.
(c) A plot of the Fe-Ch peak position, marked by a blue arrow in (a), as a function of $z$.
Two dashed lines are guides to the eye, highlighting the two-regimes behavior.
}
\end{figure}
\begin{figure}[tbp]
\includegraphics[width=1.0\columnwidth]{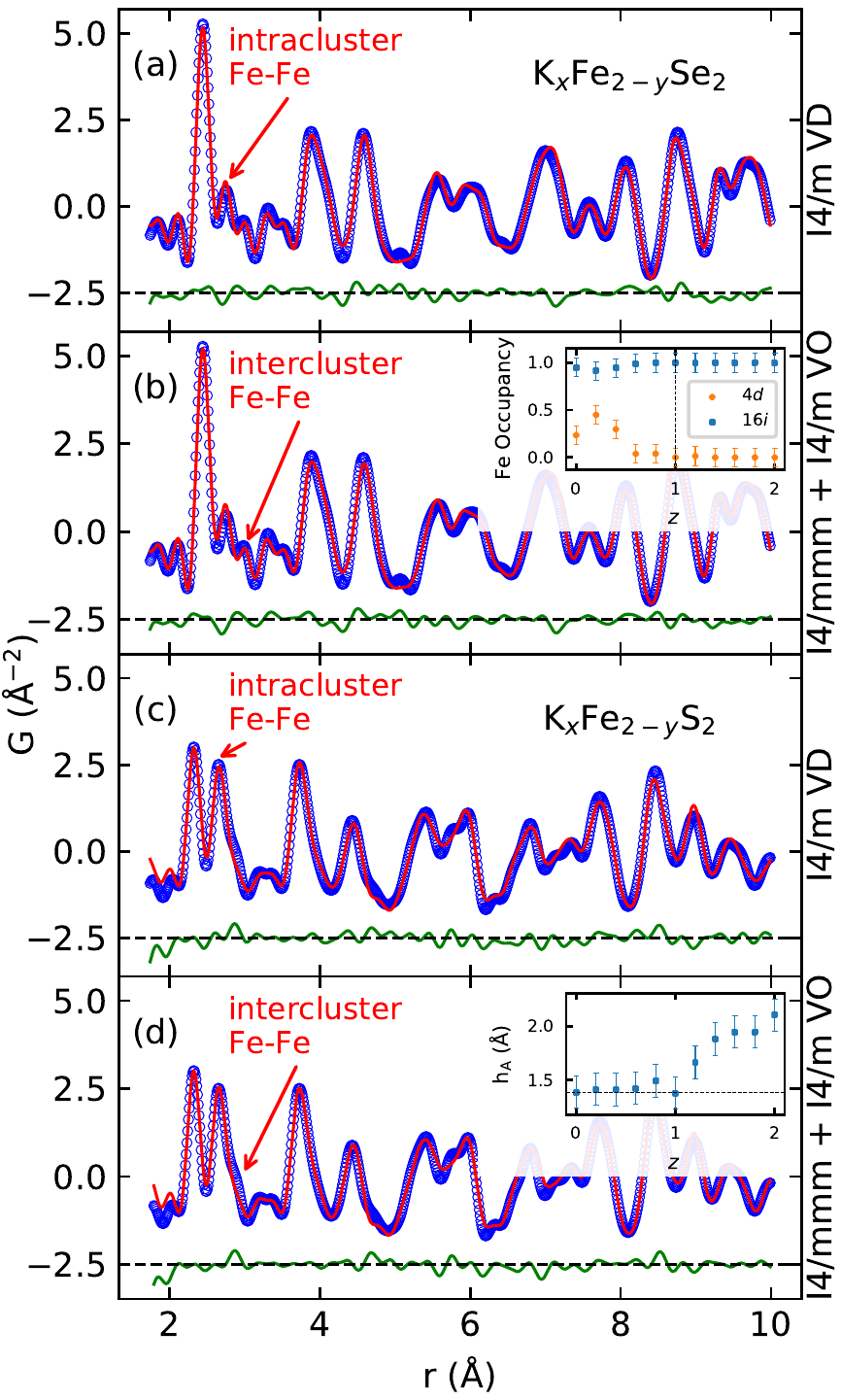}
\caption{\label{fig:fig8} (Color online) Fits of the structural models to the PDF data.
The fit to the observed K$_{x}$Fe$_{2-y}$Se$_{2}$ PDF data using either: (a) a single-phase $I4/m$ vacancy-disordered (VD) model or (b) a two-phase model with $I4/m$ VO and $I4/mmm$ vacancy-free components.
Similar one- and two-phase fits for  K$_{x}$Fe$_{2- y}$S$_{2}$ are shown in (c) and (d), respectively, for the K$_{x}$Fe$_{2- y}$S$_{2}$ PDF data.
In both compositions, single- and two-phase models yielded fits of similar quality: $R_w = 0.12$ and $R_w = 0.15$ for K$_{x}$Fe$_{2- y}$Se$_{2}$ and K$_{x}$Fe$_{2- y}$S$_{2}$, respectively.
In various panels arrows point to intra-cluster and inter-cluster Fe-Fe PDF peaks, as indicated.
See text and Fig.~\ref{fig:fig1}e for details.
In all cases, fits were over the displayed $r$-range ($1.5 \leq r \leq 10$~\AA).
PDF data and fits are shown with open circles and solid red lines, respectively, and the difference curves (solid green lines) are shown offset vertically for clarity.
The inset in (b) summarizes the single-phase PDF refinement-derived evolution with sulfur content of Fe occupancy of the two distinct Fe sites (4$d$ and 16$i$) in the $I4/m$ VD model.
Inset in (d) shows evolution of chalcogen anion height $h_A$ with sulfur content as obtained from the $I4/mmm$ vacancy-free component of the two-phase PDF model.
A horizontal dashed line represents the reported optimal anion height for superconductivity~\cite{mizuguchi_anion_2010}.}
\end{figure}
Fitting the local structure region of the observed PDFs ($1.75 \leq r \leq 10$~\AA) with a two-phase model incorporating both the VO $I4/m$ and the vacancy-free $I4/mmm$ phases yielded fits which were equivalent in quality to those done with a single-phase VD $I4/m$ model, where the occupancy of Fe on the two distinct crystallographic sites is allowed to vary.
Examples of these fits for the S- and Se-end-members are shown in Fig.~\ref{fig:fig8}.
\ins{In these two-phase models, all samples across the series possess the minority $I4/mmm$ phase with an average refined volume percentage of \textit{ca.} 8 \%.}

Importantly, only the lower symmetry $I4/m$ model supports the existence of two unique Fe-Fe correlation peaks representing inter- and intra-cluster Fe-Fe correlations (see Fig.~\ref{fig:fig1}e and Fig.~\ref{fig:fig8}), and these two unique Fe-Fe correlation peaks were observed throughout the entire composition range (see Fig.~\ref{fig:fig7}a).
Additionally, the absence of well-defined sharp (110) reflections (Fig.~\ref{fig:fig2}) indicates that perfect vacancy order cannot be present, \ins{and thus the presence of a $I4/m$ model  with exactly zero Fe 4$d$ site-occupancy is inconsistent with the diffraction data.}
\ins{The single-phase VD $I4/m$ model is also a simpler model with fewer free parameters compared to the two-phase description.}
For these reasons, the single-phase VD $I4/m$ model was preferred and utilized here.

While single-phase fits using a VD $I4/m$ model may seem \rout{inadequate}\ins{inappropriate}, as other earlier studies have found evidence for phase separation, it is important to note that the PDF analysis presented here considers only very local (less than 1~nm) structural features.
Our small-box modeling approach, shown in Fig.~\ref{fig:fig8}, suggest\instwo{s} that over sub-nm length scales the $I4/m$-VD model and $I4/m$-VO+$I4/mmm$ model are equivalent descriptions in terms of explaining the observed data.
\ins{It is important to highlight that this single phase description is not intended to contest the accepted phase separation picture of this system.}
It is possible and perhaps even likely that local correlated vacancy order may manifest as phase separation over longer length scales, but our results show that \rout{this is not discernible}\ins{the two (phase separation and a single phase with partial vacancy site occupancy) are both equivalent descriptions of} the local structure when using small-box fitting approaches.

The inset in Fig.~\ref{fig:fig8}b shows the refined occupancy of the two symmetry-distinct Fe sites, at 4$d$ and 16$i$ extracted from the single-phase $I4/m$ VD PDF fits.
\ins{This Fe site occupancy allows for the presence of both vacancy-free domains, attributed to the superconducting behavior, as well as VO AF domains.
Any refinement to a non-zero Fe 4$d$ site occupancy suggests that both domains are present.
The refined 4$d$ site occupancy then acts as an estimate of the fraction of the vacancy-free phase.}

As sulfur is substituted for selenium, the 4$d$ site moves from partially occupied to completely empty, whereas the 16$i$ site moves from partially vacant to completely occupied.
The situation where the 4$d$ Fe is completely vacant and the 16$i$ is completely occupied represents the VO phase, associated with suppression of superconductivity~\cite{wu_overview_2015}, whereas a partial occupancy suggests imperfect vacancy ordering, \ins{or the presence of both vacancy-free and VO domains in a single coherent lattice.}
Thus, we note a cross-over behavior in the vacancy distribution \ins{or phase separation} of the local structure, from partial to nearly ideal ordering.
According to our PDF fits, this transformation occurs locally in this system at $z=1.0$, or about 50~\% sulfur substitution at 5~K.

The inset in Fig.~\ref{fig:fig8}d shows the refined anion height $h_A$ in presumed the superconducting $I4/mmm$ phase from the two phase fits. The horizontal dashed line represents the reported optimal anion height for superconductivity~\cite{mizuguchi_anion_2010}. 
For $z\le1.0$, $h_A$ is constant and adopts the ideal value, whereas for $z>1.0$, $h_A$ increases beyond the ideal value.
The evolution of anion height across the series corroborates the Mizuguchi rule~\cite{mizuguchi_anion_2010} and parallels the observed suppression of superconductivity for $z>1.0$,.

\begin{figure*}[tb]
\includegraphics[width=1.0\textwidth]{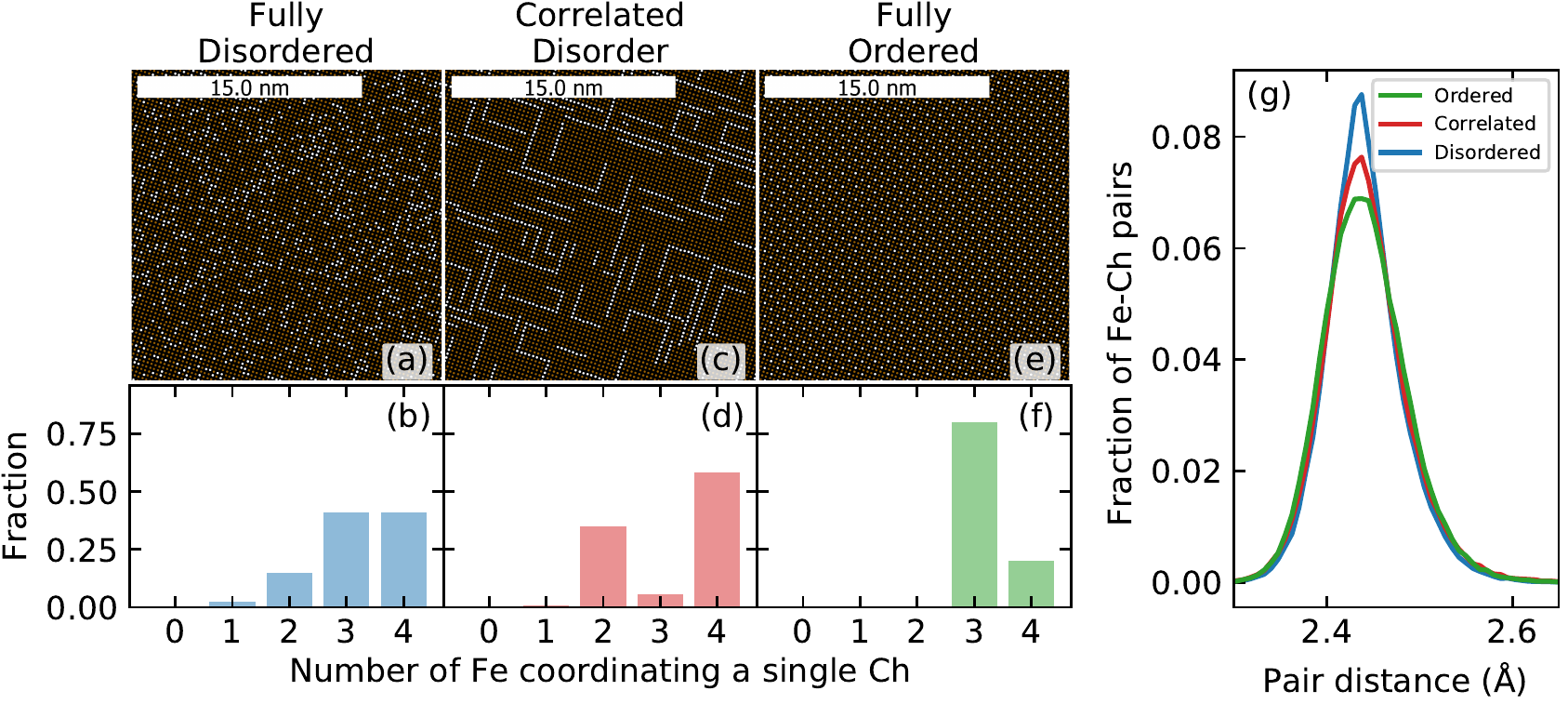}
\caption{\label{fig:fig9} (Color online) The effect of vacancy ordering on the local FeCh$_4$ structural unit.
(a) An atomistic model (viewed down the $c$-axis) featuring uniform vacancy (white dots) disorder in the iron (brown dots) sub-lattice.
(b) A histogram enumerating the total number of chalcogen species (expressed as a fraction) coordinated by either 0, 1, 2, 3, or 4 Fe in the atomic configuration represented in (a).
(c) Similar to (a), except the vacancies show correlated disorder, with an increased likelihood of like neighbors in the first coordination shell and unlike neighbors in the second coordination shell.
(d) Similar to (b) except pertaining to the atomic configuration in (c).
(e) Similar to (a), except the vacancies are fully ordered over the Fe sub-lattice.
(f) Similar to (b) except pertaining to the atomic configuration in (e).
(g) The distribution of Fe-Ch pair distances following energy minimization in the presence of a Lennard-Jones (LJ) potential between Fe-Ch nearest-neighbors, shown for the vacancy distributions represented by the atomistic model in (a), (c), and (e).
LJ potential parameters were identical between the each atomic configuration, the differences in the pair distances distributions are a result of the vacancy configuration, only.}
\end{figure*}

The results presented so far all demonstrate a two-regime behavior, and this is consistent with the electronic  phase diagram.
Electrical transport measurements suggest that $T_c$ is appreciably suppressed, beginning at $z = 1.0$, with no evidence of superconductivity above $z \approx 1.5$~\cite{lei_phase_2011}.
The structure function $F(Q)$ shows high-$Q$ oscillations which decrease in magnitude and frequency up to $z = 1.6$ and then recover for $z > 1.6$ (Figs.~\ref{fig:fig5} and~\ref{fig:fig6}).
The frequency and magnitude of these $F(Q)$ oscillations correlate with the position and relative sharpness, respectively, of the Fe-Ch NN peak as observed in the PDFs of the \kfeses\ series at 5~K.
Furthermore, this Fe-Ch NN peak disobeys Vegard's law up to about $z = 1.0$ (Fig.~\ref{fig:fig7}c) as shown by both the $F(Q)$ and $G(r)$ analysis.

While each aspect of our analysis highlights that the \kfeses\ series exhibits a two-regime behavior at 5~K, it is not immediately apparent how each individual element of this analysis \instwo{is} related to the others.
The exaggerated sharpness of the Fe-Ch NN peak (relative to higher-$r$ PDF peaks) in the selenium rich regime signifies that the Fe-Ch bond length distribution (BLD) is significantly narrower compared to Fe-Ch BLD in the sulfur rich regime~\cite{mangelis_nanoscale_2018}.
This Fe-Ch BLD in K$_{x}$Fe$_{2- y}$Se$_{2}$ (Fig.~\ref{fig:fig5}b) is also much sharper even than that seen in FeSe, observed under similar conditions (Fig.~\ref{fig:fig5}f).
This is intriguing, as the Fe-Ch NN correlations in both K$_{x}$Fe$_{2- y}$Se$_{2}$ and FeSe are due to the same structural motif of FeSe$_4$ edge-shared tetrahedra.
Thus, the observed differences in the Fe-Se BLDs between K$_{x}$Fe$_{2- y}$Se$_{2}$ and FeSe cannot be due to chalcogen size mismatch.
It is then also possible that something other than simple steric differences between sulfur and selenium contribute to the differences in relative Fe-Ch BLDs (peak sharpness) across the \kfeses\ series.

Notably, the results presented so far further suggest that the two regime behavior is marked by a crossover from a \instwo{ correlated disorder state} to more complete vacancy-order within the local structure.
While it is expected that vacancy ordering impacts the overall disorder of the structure, it is also possible that it directly impacts the Fe-Ch BLD as observed through the first PDF peak.

To explore this aspect further, we considered three large atomistic models with different local vacancy distributions but similar average structure. The first was built with a uniformly disordered arrangement of vacancies (Fig.~\ref{fig:fig9}a), the second with correlated vacancy disorder (Fig.~\ref{fig:fig9}c), and the third with a completely ordered arrangement of vacancies (Fig.~\ref{fig:fig9}e).
Importantly, these three configurations were created with identical overall vacancy concentrations (20~\%\ins{, as suggested by previous EDX studies on these samples}~\cite{lei_phase_2011}) using the same space-group ($I4/m$) and unit cell parameters.


If each of these large configurations are folded into a single unit cell, they would appear nearly identical, with the primary difference being the occupancy of the distinct Fe sites in the $I4/m$ space group.
It is clear however, that the local structure of each is quite distinct.
This local vs. average discrepancy highlights the shortcomings of traditional crystallography in describing such materials~\cite{Welberry1985,proffen_analysis_2000, Keen2015}, and reiterates our earlier observations of pervasive anisotropic broadening of the (110) diffraction peak as demonstrated in Fig.~\ref{fig:fig2}. 

In these atomic configurations, each chalcogen species can be coordinated by either four Fe, four vacancies, or permutations of both Fe and vacancies.
For each atomistic model, we quantified the total number of each of the five possible configurations.
These results are presented in Fig.~\ref{fig:fig9}b,d,f in form of normalized histograms showing the distribution of Ch with a specific number of Fe neighbors for the VD, correlated-disorder, and VO atomistic models, respectively.

For the VO case, the results are as expected, 20~\% of chalcogens are fully coordinated by four Fe, whereas 80~\% of chalcogens are coordinated by three Fe and a single vacancy, and no chalcogens are coordinated by two or more vacancies.

The VD case is more interesting: a significant fraction of chalcogens ($\sim20$~\%) are severely under-coordinated, with two or more missing Fe NN.
Moreover, and importantly, the percentage of fully coordinated chalcogens nearly doubles compared to the VO case.
These vacancy-free configurations can be associated with the presumed superconducting $I4/mmm$ phase~\cite{lei_phase_2011}.
Surprisingly, and somewhat counter-intuitively, the increase in fully coordinated chalcogens is a result of disordering the vacancies in this system.
The atomic configuration exhibiting correlated disorder however shows the greatest prevalence of fully coordinated chalcogens, notably the vacancy-free domains represent more than $50$~\%, greater than the percolation threshold in a square lattice.

It is interesting to note that the atomic arrangement depicted in Fig.~\ref{fig:fig9}c shows a distinct pattern, where linear stripes of vacancies meet at right angles and are arranged around vacancy-free domains. 
This mirrors the striking microstructures observed in many cases in the Se-end-member (K$_{x}$Fe$_{2-y}$Se$_{2}$) system~\cite{ding_influence_2013, PhysRevB.91.064513, liu_formation_2016}.

If the three configurations exemplified by Fig.~\ref{fig:fig9}a,c,e are energy minimized in the presence of an \textit{identical} Lennard-Jones potential between Fe and Ch NNs, as outlined fully in in Appendix~\ref{sec:appe}, the resulting Fe-Ch BLDs, shown in Fig.~\ref{fig:fig9}g, are \textit{different}.
Specifically, the Fe-Ch BLD of the energy minimized VD configuration is noticeably sharper than the VO configuration, whereas the configuration showing correlated disorder represents an intermediate case.
It is important to note that the energy minimization did not involve site swapping or vacancy rearrangement, and as such the resulting atomistic models are indistinguishable from those shown in Fig.~\ref{fig:fig9}a,c,e.
Our finding is quite surprising, given that the sharpest distribution originates from a system with the most overall disorder.

Our simulations demonstrates that the relative arrangement of vacancies has an intrinsic impact on the Fe-Ch BLD, even in the presence of identical pair-wise interactions.
This idea makes intuitive sense, as the VD state shows a greater prevalence of severely undercoordinated chalcogen species.
Using the language of ``correlated motion'' of nearest neighbors, this could imply that these severely under-coordinated chalcogens will form a more rigid bond with the fewer Fe species they are coordinated by, leading to a relatively narrower Fe-Ch BLD.

The result also offers an explanation of the two-regime behavior.
Specifically, for the selenium-rich cases the local structure may show correlated or imperfect vacancy order, characterized by domains with no vacancies coexisting in the same lattice with domains showing ordered vacancies. 
Conversely, in the sulfur-rich cases, a vacancy redistribution may create more idealized ordering.
This is further supported by the increased prevalence of the longer inter-layer spacing towards the sulfur end of the phase diagram, as shown in Fig.~\ref{fig:fig2} and discussed in Fig.~\ref{fig:fig3}b. 
This longer inter-layer spacing has been linked in previous work on the Se-end-member to localized domains with larger than average potassium concentrations~\cite{PhysRevB.91.064513}.
This same study has further shown that Fe-vacancy ordering is linked closely to superconductivity and the distribution of potassium~\cite{PhysRevB.91.064513}.

This effect manifests in $F(Q)$ through suppression of Bragg peaks at high-$Q$ (Fig.~\ref{fig:fig6}), leading to the observation of a strong sinusoidal intensity oscillation.
In $G(r)$ this oscillation manifests as a relatively sharp first peak followed by broader high-$r$ peaks (Fig.~\ref{fig:fig7}), corresponding to a sharper Fe-Ch BLD.

It is likely that the local structure of each sample sits somewhere between the fully VD and fully ordered case, represented by Fig.~\ref{fig:fig9}a and Fig.~\ref{fig:fig9}e respectively.
To further quantify the degree of correlated disorder, we have conducted structure-model independent analysis of the first three measured PDF peaks across the series.
These peaks correspond to Fe-Ch, Fe-Fe intra- and inter-cluster correlations, in order of increasing $r$.

\begin{figure}[tbp]
\includegraphics[width=1.0\columnwidth]{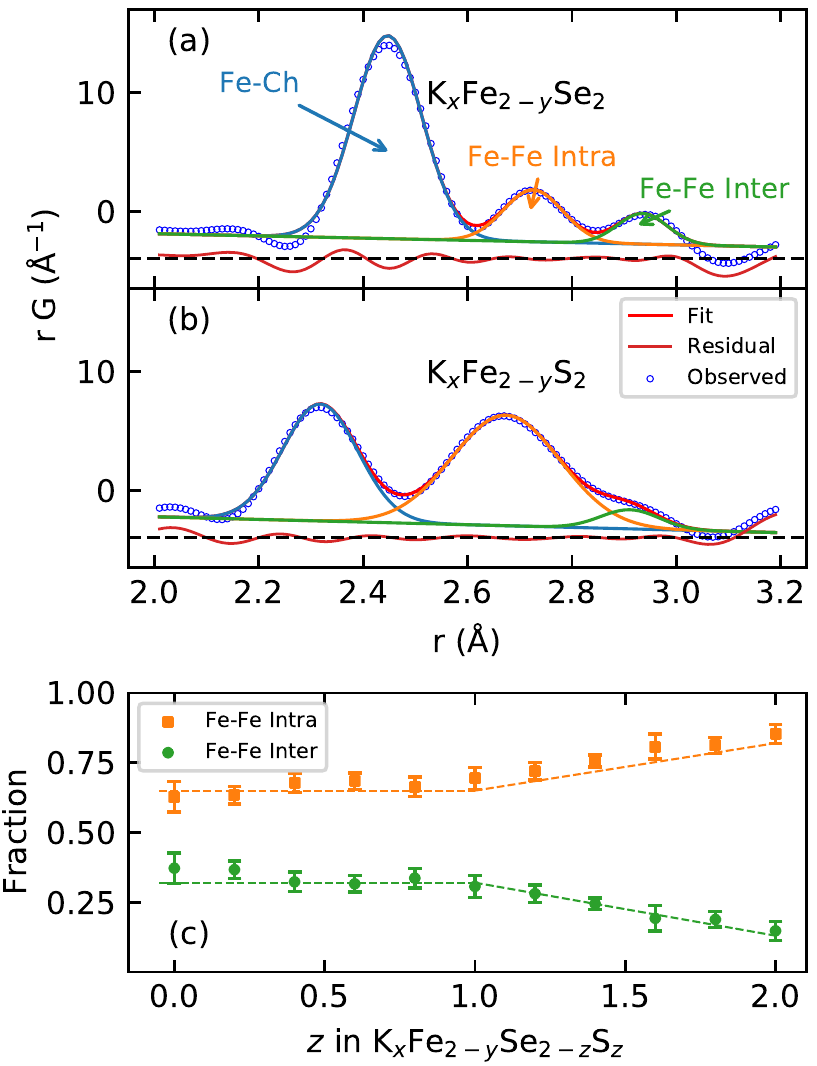}
\caption{\label{fig:fig10} (Color online) Evolution of the local structure with sulfur content.
(a) K$_{x}$Fe$_{2-y}$Se$_{2}$ radial distribution function showing Fe-Ch and distinct Fe-Fe pair distances (see text).
(b) Similar to (a) except with K$_{x}$Fe$_{2-y}$S$_{2}$, highlighting the relative increase in prevalence of intra-cluster Fe-Fe pairs.
(c) Quantification of the prevalence of different inter- and intra-cluster Fe-Fe pairs as a function of sulfur content $z$.
(d) Quantification of the inter- and intra-cluster Fe-Fe pair distances as a function of sulfur content $z$.}
\end{figure}

Fig.~\ref{fig:fig10}a,b shows example fits of Gaussian profiles to the radial pair distribution function $r G(r)$ for the selenium and sulfur end-members, respectively. 
As the Fe-Fe intra- and inter-cluster peaks are a result only of Fe-Fe correlations, their mutual \textit{relative} contributions to the PDF are not impacted by the overall system chemistry, and their area normalized to the area of the pair give an indication of the prevalence of the distinct Fe-Fe environments.
In Fig.~\ref{fig:fig10}c we show the relative fraction of Fe-Fe intra- and inter-cluster pairs as obtained from this analysis.
We see that for the selenium-rich phases up to $z=1.0$ the relative abundance of each type of Fe-Fe pair is largely unchanged.
In the sulfur-rich portion when $z>1.0$ there is a monotonic increase in the fraction of intra-cluster pairs at the expense of inter-cluster pairs.
This is consistent with vacancy redistribution and an increase in vacancy ordering when moving from Se-rich toward S-rich end of the phase diagram, as the fully ordered configuration should maximize the fraction of intra-cluster Fe-Fe pairs.
This analysis further supports the two regime behavior of this system in phase-space.

\begin{figure}[tbp]
\includegraphics[width=1.0\columnwidth]{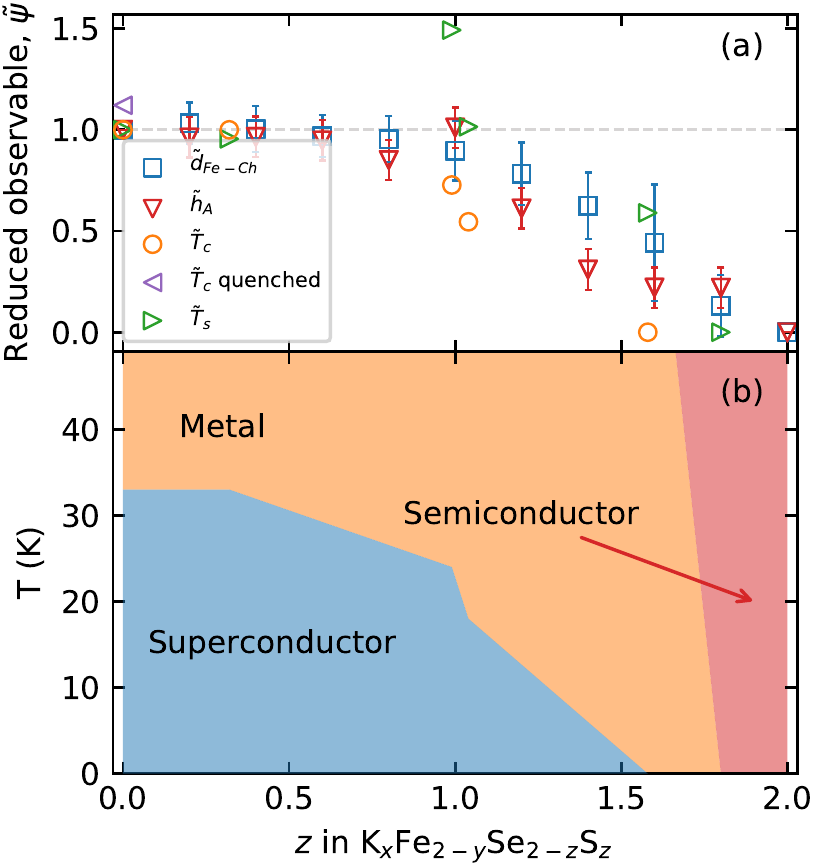}
\caption{\label{fig:fig12} (Color online) Qualitative comparison of PDF derived local structural parameters and electronic properties of \kfeses.
(a) Reduced observables, representing electronic and structural quantities rescaled to a fraction of their values at the end-members.
\ins{By definition (see text) these reduced observables are one at $z=0$ and zero at $z=2$.}
(b) The previously reported~\cite{lei_phase_2011} electronic phase diagram of \kfeses\ \rangez\  as a function of temperature $T$ and composition $z$, \ins{obtained by resisitivity and dc magnetic susceptibility measurements using a Quantum Design PPMS-9 and MPMS-XL5 on single crystal samples prior to grinding.} \rout{reflecting electrical transport and magnetization measurements.}}
\end{figure}

Finally, we explore the correlation between the measures of local structure outlined here to the electronic properties in this system in the context of ``reduced observables'' computed as a function of sulfur content, $\tilde{\psi}_{red}(z)$.
We define a reduced observable as the observed quantity at a given composition $z$ normalized by the same observed quantities at the composition end-members ($z=0$ and $z=2$) such that $\tilde{\psi}_{red}(z) = \frac{ \psi(z)-\psi(z=2)}{\psi(z=0)-\psi(z=2)}$.
By design, $\tilde{\psi}_{red}(0)=1$ and $\tilde{\psi}_{red}(2)=0$.

\rout{Our}Results for the superconducting temperature $\tilde{T}_c$ \ins{ and metal-insulator transition temperature $\tilde{T}_s$, reported previously}~\cite{lei_phase_2011} \ins{and} the Fe-Ch distance \ins{$\tilde{d}_{Fe-Ch}$}, \ins{as well as} the anion height $\tilde{h}_A$ \ins{derived from the present PDF analysis,} \rout{,and metal-insulator transition temperature $T_s$} are plotted in Fig.~\ref{fig:fig12}a.
Also included for reference is the $T_c$ of a quenched K$_{x}$Fe$_{2-y}$Se$_{2}$ sample.
Importantly, $T_c$ for quenched sample is higher than for as-prepared, in line with the idea that stacking disorder and potassium distribution play an important role for superconducting properties of this system~\cite{PhysRevB.84.212502}.

The results are striking, as each of these reduced quantities largely follow the same trend, remaining essentially constant up to $z=1$, then decreasing steadily in the range $1<z <2$.
This serves to emphasize the apparent connection between the sub-nanometer atomic structure at 5~K, as studied here, and the electronic properties (see e.g. Fig.~\ref{fig:fig12}b and \ins{Ref.}~\cite{lei_phase_2011}), which invites further complementary  experimental and theoretical considerations. 
\section{Conclusions}
In summary, the local atomic structure and structural disorder have been characterized across the phase diagram of \kfeses\ \rangez\ system, displaying both superconductivity and magnetism, by means of neutron total scattering and associated atomic PDF analysis of 5~K data.
Using various structure-model independent and model dependent approaches we find a high level of structural complexity on the nanometer length-scale.
The analysis highlights the presence of considerable disorder in the $c$-axis stacking of the Fe$_2$Se$_{2-z}$S$_{z}$ slabs, which has a non-turbostratic character, and can be suppressed by thermal quenching.
\ins{PDF fitting suggests that over sub-nanometer length scales a simple two-phase physical mixture description (VO and vacancy-free) is equivalent to a single phase description allowing for partial Fe occupancy of the vacancy site.
Within this description, a non-zero Fe occupancy of the $4d$ site indicates the presence of vacancy free superconducting domains.}
Complementary aspects of the analysis reveal a crossover like behavior, with an onset at $z\approx1$, from predominantly \instwo{correlated disorder} state towards the selenium end of the phase diagram, to a more ordered vacancy distribution closer to the sulfur end of the phase diagram.
Through simulations based on large scale atomistic models we demonstrate that the distribution of vacancies can significantly modify the NN bond length distribution, observably affecting the NN environment, as corroborated by the features in the data.
\ins{The correlation of electronic properties with the Fe vacancy distribution may imply that the disorder of Fe vacancies in the local structure creates `domains' resembling Fe-vacancy-free superconducting $I4/mmm$ phase, which may be suppressed with the correlated disorder to order crossover across the series.}
\ins{This could imply that the SC bearing vacancy-free phase is suppressed with S substitution, although our analysis does not directly probe this aspect.}
\rout{The evolution of the local structure with sulfur content displays apparent correlation with the changes seen in the electronic state, emphasizing the importance of the local structure for the observed electronic properties.}
\ins{The local structure observations and their correlation with electronic properties presented here add to the nuanced picture of the role of atomic structure in this system, and should help to inform future experimental and theoretical work seeking to understand the mechanisms behind these correlations.}\rout{The results demonstrate the direct impact of the Fe-vacancy distribution on the local structure in this system, and reinforce the idea of its critical influence on superconductivity.}

%
%
\section{acknowledgments}
Work at Brookhaven National Laboratory was supported by U.S. Department of Energy, Office of Science, Office of Basic Energy Sciences (DOE-BES) under contract DE-SC0012704.
Alexandros Lappas acknowledges support by the U.S. Office of Naval Research Global, NICOP grant award No. N62909-17-1-2126.
This research used resources at the Spallation Neutron Source, a U.S. Department of Energy Office of Science User Facility operated by the Oak Ridge National Laboratory.
Notice this paper has been co-authored by UT-Battelle, LLC under Contract No. DE-AC05-00OR22725 with the U.S.
Department of Energy. 
\instwo{The United States Government retains and the publisher, by accepting the article for publication, acknowledges that the United States Government retains a nonexclusive, paid-up, irrevocable, world wide license to publish or reproduce the published form of this paper, or allow others to do so, for United States Government purposes.
The Department of Energy will provide public access to these results of federally sponsored research in accordance with the DOE Public Access Plan (http://energy.gov/downloads/doepublic-access-plan)}

\appendix
\section{Neutron total scattering}
\label{sec:appa}
Experiments were performed using powder samples, 0.5~g of each, equally spaced across the Se/S compositional space $(\Delta z = 0.2)$, loaded into 6~mm diameter extruded vanadium containers under inert atmosphere and sealed.
Each sample was mounted in the diffractometer equipped with Orange cryostat.
The instrument was calibrated using diamond powder standard.
Powder diffraction data were collected after thorough equilibration at 5~K for 2~h of total counting time for each sample.
Neutron PDFs were obtained via Sine Fourier transforms over a range from \qmin\ = 0.5~\RAA\ to \qmax\ = 26~\RAA.

\section{Structure-model independent analysis}
\label{sec:appb}
Fitting of PDF data for decomposition into constituent peaks was done with the fityk v 1.3.1 software package~\cite{Wojdyr2010} using Gaussian functions and a linear baseline function where the intercept term was fixed at zero.
The $F(Q)$ data for each sample were fitted with a damped Sine function after subtracting a fitted linear background.

\section{Structural models}
\label{sec:appc}
 Within the system, two possible structural models have been put forward.
The simplest model, with space group $I4/mmm$, consists of FeCh (Ch = Se/S) slabs featuring a Fe square planar sublattice (Fig.~\ref{fig:fig1}a-c).
Each Fe is coordinated by four Ch creating layers of edge-shared FeCh$_4$-tetrahedra, stacked along the c lattice direction and interleaved with K species equidistant from each layer.
In this model, the asymmetric unit contains a single Fe at 4$d$ $(0.0, 0.5, 0.25)$, Ch at 4$e$ $(0.0, 0.0, z)$, and K at 2$a$ $(0.0, 0.0, 0.0)$.
The relatively simple structure, with only one unique Fe site, gives little flexibility for handling vacancies on the Fe sub-lattice, which are common in this system~\cite{carr_structure_2014,ding_influence_2013,wu_overview_2015}.

A more complicated model, with space group $I4/m$, is related to the higher symmetry $I4/mmm$ model through a rotation in the $a-b$ plane and a $\sqrt{2}$ increase in the $a = b$ lattice parameters (Fig.~\ref{fig:fig1}d,e).
The increase in unit cell size doubles the number of atoms in the asymmetric unit, with two symmetry-distinct Fe at 4$d$ $(0.0, 0.5, 0.25)$ and 16$i$ $(x, y, z)$, Ch at 4$e$ $(0.0, 0.0, z)$ and 16$i$ $(x, y, z)$, and K at 2$b$ $(0.0, 0.0, 0.5)$ and 8$h$ $(x, y, 0.5)$.
Importantly, within both these symmetries the FeCh$_4$ tetrahedra of adjacent layers are translated in the x and y direction by half unit cell lengths.
This can be contrasted with the related FeCh-11 group of superconductors, where adjacent layers of FeCh$_4$-tetrahedra contain no such relative $x$ and $y$ translation.

\section{Small-box refinements}
\label{sec:appd}
The experimental PDF data were fit with the structural models described over a 10~\AA\ range using the \pdfgui\ software~\cite{farro;jpcm07}.
Over wider $r$-ranges the experimental PDF is dominated by the inter-layer correlations, and our analysis reveals that these are affected by appreciable stacking disorder, which is highly non-trivial to handle using conventional PDF approaches and is beyond the scope of this study.
Two alternative modeling approaches were attempted in sub nanometer regime.
The first approach used a mixture of vacancy-free $I4/mmm$ and fully Fe VO $I4/m$ phase components.
The second approach utilized only a VD version of $I4/m$.
The two approaches were found to yield effectively identical fit qualities and comparable descriptions of the underlying structure~\cite{mangelis_nanoscale_2018}.
The VD model with $I4/m$ symmetry, systematically applied to all data, used a total of 19 fitting parameters.
These included the unit cell parameters $(a=b \ne c)$, a scale factor, and a correlated motion parameter $\delta_1$~\cite{jeong;jpc99}.
Further, the fractional coordinates were refined according to the space group constraints (9 parameters), and the atomic displacement parameters (ADP) were set to be isotropic ($u_{11} = u_{22} = u_{33}$) and identical for all atomic species of the same type (3 parameters).
The occupancies of the two symmetry-distinct Fe crystallographic sites (4$d$ and 16$i$) were allowed to vary separately, while the occupancies of potassium atoms in 2$b$ and 8$h$ sites were constrained to be equal because of relative insensitivity of the neutron PDF to potassium (3 occupancy parameters).
The occupancies of Ch species were fixed at their nominal values as refinements did not suggest chalcogen sub-lattice ordering.
Refinements were done sequentially, such that for any given composition the PDF refinement was initialized using the converged model of the previous composition.
Additionally, appropriately constrained two-phase fits were used to track the evolution of the anion height estimated from the $I4/mmm$ vacancy-free component.

\section{Large-box simulations}
\label{sec:appe}
Simulated PDFs were computed with aid of the DISCUS v 5.30.0 software package~\cite{Proffen1997} using a model with space group $I4/m$.
The global concentration of Fe was fixed at the nominal concentration, with full occupancy of the 16$i$ site and full vacancy occupancy of the 4$d$ site.
The global concentration of K was fixed at the nominal concentration, spread uniformly across the 2$b$ and 8$h$ sites.
For stacking fault simulations, the stack module of the DISCUS software package was used, with a layer supercell of $2\times2$ unit cells and total thickness of 1,500 layers.
The powder pattern was computed using the Fast Fourier method in DISCUS, with a 0.001 reciprocal length unit (r.l.u) mesh for powder integration.
For simulations on the impact of vacancy disorder, a supercell of $50\times50\times1$ unit cells was used.
Atoms were initialized with a random thermal displacement such that their mean squared displacement across the whole supercell was consistent with the ADP parameters refined from the \kfeses\ PDF data.
Following this, atomic displacement vectors were swapped between atoms of like identity to minimize the total energy $E_{LJ}$ composed of a pairwise Lennard Jones (LJ) potential between NN Fe-Ch pairs, specifically 
\begin{equation}
E_{LJ}=\sum_{i}^{}\sum_{n\neq i}^{}\left[\frac{A}{d_{in}^{12}}-\frac{B}{d_{in}^{6}} \right ]
\end{equation}
over all atoms $i,n$, with 
\begin{equation}
A=D\frac{1}{2}\tau^{12}_{in} \textrm{ and } B =2 D \tau^{6}_{in}.
\end{equation}
The LJ potential was constructed such that the equilibrium NN Fe-Ch bond distance was that refined from the \kfeses\ PDF data ($\tau_{in} = 2.44$~\AA) with potential depth $D = 100$.
Swaps were always accepted if they decreased the total energy, and conditionally accepted if they increased the total energy, with probability $p=\exp{(- \Delta E/kT)}$, where $\Delta E$ is the change in energy associated with the swap, $k$ is the Boltzmann constant, and $T$ is the temperature, in this case 5~K.
The total number of swaps was fixed at 100 times the number of atoms in the supercell.

\bibliography{billinge-group,abb-billinge-group,everyone,19_rk_122_FeSCs}
\bibliographystyle{apsrev}

\end{document}